\documentclass[doublecol]{epl2}

\usepackage{graphicx,amsmath}

\newcommand{\figwidth}{0.95\columnwidth}
\newcommand{\eq}[1]{Eq.(\ref{#1})}
\newcommand{\fig}[1]{Fig.~\ref{#1}}
\newcommand{\avg}[1]{ {\langle #1 \rangle} }
\newcommand{\ahum}[1]{``#1''}
\newcommand{\zbcr}{z_{B,\rm cr}}
\newcommand{\zcr}{z_{\rm cr}}

\newcommand{\beq}{\begin{equation}}
\newcommand{\eeq}{\end{equation}}
\newcommand{\bea}{\begin{eqnarray}}
\newcommand{\eea}{\end{eqnarray}}

\title{Restricted orientation \ahum{liquid crystal} in two 
dimensions: \\ isotropic--nematic transition or liquid--gas (?)}

\shorttitle{Restricted orientation liquid crystal in two dimensions}

\author{T. Fischer \and R. L. C. Vink}

\shortauthor{T. Fischer and R. L. C. Vink}

\institute{Institute of Theoretical Physics, Georg-August-Universit\"at 
G\"ottingen, Friedrich-Hund-Platz~1, \\ 37077 G\"ottingen, Germany}


\abstract{We present Monte Carlo simulation results of the two-dimensional 
Zwanzig fluid, which consists of hard line segments which may orient either 
horizontally or vertically. At a certain critical fugacity, we observe a phase 
transition with a two-dimensional Ising critical point. Above the transition 
point, the system is in an ordered state, with the majority of particles being 
either horizontally or vertically aligned. In contrast to previous work, we 
identify the transition as being of the liquid-gas type, as opposed to 
isotropic-to-nematic. This interpretation naturally accounts for the observed 
Ising critical behavior. Furthermore, when the Zwanzig fluid is extended to more 
allowed particle orientations, we argue that in some cases the symmetry of a 
$q$-state Potts model with $q>2$ arises. This observation is used to interpret a 
number of previous results.}

\begin{document}

\maketitle

\section{Introduction}

In a seminal paper \cite{onsager:1949}, Onsager demonstrated that infinitely 
slender rods in three dimensions undergo a first-order isotropic-to-nematic (IN) 
transition. In the nematic phase, there is long-ranged alignment of the 
particles, while in the isotropic phase the particle orientations are 
essentially random. In contrast, in two dimensions, long-ranged nematic order is 
generally absent. For a certain class of liquid crystal pair potentials, the 
absence of nematic order can be proved rigorously \cite{physreva.4.675}, while 
simulations using different potentials also indicate its absence in the 
thermodynamic limit \cite{physreva.31.1776,bates.frenkel:2000}. Of course, these 
results do not imply that there can be no phase transition in two-dimensional 
(2D) liquid crystals, but rather that any such transition does not lead to 
nematic order.

Interestingly, a number of papers have appeared recently \cite{ghosh.dhar:2007, 
matoz-fernandez.linares.ea:2008, linares.roma.ea:2008, matoz-fernandez:214902} 
in which the IN transition was studied in two dimensions. The transition was 
shown to belong to the universality class of the 2D Ising model. In accordance 
with the Ising model, this implies the formation of a finite nematic order 
parameter above the critical density, which seems to contrast the results of 
\cite{physreva.31.1776,bates.frenkel:2000}, where nematic order was found to 
vanish in the thermodynamic limit. The results of \cite{ghosh.dhar:2007, 
matoz-fernandez.linares.ea:2008, linares.roma.ea:2008, matoz-fernandez:214902} 
thus raise a number of questions. First of all, how can we understand the 
formation of finite nematic order in these 2D systems and, secondly, what is the 
origin of the Ising critical point? In this work, these questions will be 
answered.

The absence of nematic order in many two dimensional systems is a consequence of 
the Mermin-Wagner theorem \cite{physrevlett.17.1133, physrevlett.87.137203, 
ioffe.shlosman.ea:2002}. As is well known, this theorem applies when the 
particle orientations are {\it continuous}. However, when the orientations 
become {\it discretized}, Mermin-Wagner no longer applies, and the corresponding 
phase behavior changes dramatically. A famous example of a liquid crystal model 
with discrete orientations is the Zwanzig model \cite{zwanzig:1714}, where the 
particles are treated as rigid rods. The particle positions are continuous, but 
the molecular axis may only point in mutually perpendicular directions. In two 
dimensions, this implies a system of line segments, which may either point 
horizontally or vertically. The interactions are of the excluded volume type, 
meaning that particles may not overlap with each other. A further approximation 
is to also make the particle positions discrete, i.e.~to restrict the line 
segments to the sites of a square lattice, and to let each segment occupy $k$ 
consecutive sites. This is precisely the model studied in \cite{ghosh.dhar:2007, 
matoz-fernandez.linares.ea:2008, linares.roma.ea:2008, matoz-fernandez:214902}.  
For $k=2$ one recovers the dimer model \cite{temperley.fisher:1961}, while for 
$k \to \infty$ one approaches the 2D Zwanzig model. Provided $k \geq 7$ 
\cite{matoz-fernandez:214902}, at some threshold density, one finds a transition 
to a \ahum{nematic} phase, in which most particles point either horizontally 
(``A particles'') or vertically (``B particles'').

However, is this transition truly an IN transition? In this model, there is 
symmetry under the exchange of A and B particles. That is, given a valid 
configuration, i.e.~one without overlaps, a new valid configuration can be 
obtained by replacing each A~particle with a B~particle and vice-versa. Under 
this operation, the order parameter $|N_A - N_B|/(N_A+N_B)$ 
\cite{ghosh.dhar:2007} remains invariant, with $N_i$ the number of particles of 
type $i$. Clearly, a liquid crystal with {\it continuous} orientations cannot 
exhibit this symmetry. The observed symmetry rather resembles the particle-hole 
symmetry of the lattice gas, or the up-down symmetry in the Ising model. This 
suggests that the IN transition of \cite{ghosh.dhar:2007, 
matoz-fernandez.linares.ea:2008, linares.roma.ea:2008, matoz-fernandez:214902} 
is really an unmixing or liquid-gas transition. This also accounts for the 
observed Ising critical behavior, since it is well known that liquid-gas and 
unmixing transitions in fluids with short ranged interactions belong to this 
class.

The interpretation in terms of a liquid-gas transition is also consistent with 
the original paper by Zwanzig \cite{zwanzig:1714}. Here, it was already 
mentioned that, in two dimensions, a mapping of the Zwanzig model onto a mixture 
of green and red squares \cite{hoover:3141} is possible, whereby squares of 
different color may not overlap. This is, of course, just a variant of the 
Widom-Rowlinson mixture of spherical particles in two dimensions 
\cite{widom.rowlinson:1970}, in which the existence of a liquid-gas transition 
is not debated. Note that the origin of liquid-gas transitions in these systems 
stems from {\it depletion}. One could envision formally integrating out, say, 
the \ahum{red} species, yielding a one component fluid of \ahum{green} species, 
interacting via an effective short-ranged potential with attractive part 
\cite{johnson.gould.ea:1997}. Hence, these systems resemble simple fluids, such 
as the Lenard-Jones fluid, and are expected to yield similar phase diagrams as a 
result.

In this Letter, these ideas will be applied to the 2D Zwanzig model 
using computer simulation. We first specify the 2D Zwanzig model, and 
describe the simulation method. Next, we show that the transition in 
this model indeed corresponds to a liquid-gas transition, rather than 
IN. In particular, we demonstrate that a binodal can be constructed, 
which terminates at an Ising critical point. We also provide estimates 
for the line-tension between coexisting domains in the two-phase 
(ordered) region of the phase diagram. Finally, we present a summary and 
detailed conclusion, where we emphasize that care must be taken when 
modeling liquid crystal phase transitions using only a discrete set of 
orientations.

\section{model and simulation method}


We consider the 2D Zwanzig model, which consists of infinitesimally thin hard 
rods of unit length. Since the interactions are hard-core, temperature does not 
play a role, and factors of $k_B T$ are set to unity throughout (with $k_B$ the 
Boltzmann constant and $T$ the temperature). The rods may be aligned 
horizontally (``$A$ particles'') or vertically (``$B$ particles''). The particle 
positions are confined to a periodic 2D square of area $V$. Since the rods are 
infinitesimally thin, there is only a hard-core interaction between $A$ and $B$ 
particles (which may thus not overlap). Hence, each $A$ particle is surrounded 
by a {\it depletion zone}, which may not contain the centers of any $B$ 
particles (note that the depletion zone for this model is just the unit square). 
We consider a grand canonical simulation ensemble, with the respective chemical 
potentials $\mu$ and $\mu_B$, of $A$ and $B$ particles, being the relevant 
thermodynamic parameters; the actual number of particles in the system is a 
fluctuating quantity. The aim of the simulations is to measure the distribution 
$P(N|\mu,\mu_B)$, defined as the probability to observe a system containing $N$ 
particles of type $A$, at chemical potentials $\mu$ and $\mu_B$.


During the simulations, insertion and removal of particles are performed using a 
grand canonical cluster move \cite{vink.horbach:2004*1, vink:2004}. With equal 
probability, we attempt to insert an $A$ particle, or we attempt to remove one. 
When inserting, a single $A$ particle is tentatively placed at a random 
location in the simulation box. This will generally lead to overlap with some 
$B$ particles, say $n_B$ of them. The overlapping $B$ particles are removed from 
the box, and the resulting state is accepted with probability
\begin{eqnarray*}
 A(N \to N+1,N_B \to N_B-n_B) = \hspace{2cm} \nonumber \\
 \begin{cases}
 0 & n_B \geq \Delta \\
 {\rm min}\left[1, \frac{z V}{\Delta (N+1)} \frac{(n_B)!}{z_B^{n_B}} \right] 
 & \text{otherwise},
 \end{cases}
\end{eqnarray*}
with $\Delta$ a parameter to be specified later, and $N_B$ the total number of 
$B$ particles in the system at the beginning of the move. In the above, we have 
also introduced the respective fugacities $z = \exp(\mu)$ and $z_B = 
\exp(\mu_B)$, of $A$ and $B$ particles. During removal, one $A$ particle is 
picked randomly and deleted, and $n_B$ centers of $B$ particles are distributed 
randomly into the depletion zone of the just deleted $A$ particle, with $n_B$ a 
uniform random number $0 \leq n_B < \Delta$. If any of the inserted $B$ 
particles overlap with $A$ particles, the move is rejected, otherwise it is 
accepted with probability
\begin{eqnarray*}
 A(N \to N-1,N_B \to N_B+n_B) = \hspace{2cm} \nonumber \\
 {\rm min}\left[1, \frac{\Delta N}{z V} \frac{z_B^{n_B}}{(n_B)!} \right].
\end{eqnarray*}
The reader may verify that this algorithm fulfills detailed balance 
\cite{vink.horbach:2004*1, vink:2004}. The factorials count the number of ways 
in which $n_B$ particles can be distributed onto the unit square. Note that the 
thermal wavelength has been set to unity for clarity. The parameter $\Delta$ 
must be high enough such that the insertion of $A$ particles into a pure phase 
of $B$ particles is efficient. For the present model, a pure phase of $B$ 
particles is just an ideal gas, for which density equals fugacity. Hence, the 
depletion zone contains $z_B$ $B$ particles on average, with Poissonian 
fluctuations. Consequently, $\Delta$ should somewhat exceed this value; we found 
that $\Delta = z_B + \sqrt{z_B} + 2$ gave good results.


The phase transition in our model is characterized by a free energy functional 
featuring two minima separated by a barrier. Apart from a minus sign, the free 
energy is just the logarithm of the distribution $P(N|\mu,\mu_B)$ that we wish 
to find, and we define $W(N) \equiv \ln P(N)$. Computer simulations which 
directly sample the Boltzmann distribution are not efficient then, since these 
tend to \ahum{get stuck} in one of the minima, and rarely cross the barrier. To 
overcome this problem, we combine the grand canonical cluster move with a biased 
sampling method called successive umbrella sampling (SUS) 
\cite{virnau.muller:2004}. In SUS, $W(N)$ is obtained recursively by splitting 
the simulation into a number of windows. In the first window, the number of $A$ 
particles is allowed to fluctuate between 0 and 1, in the second window between 
1 and 2, and so forth. There is no restriction on the number of $B$ particles 
though, and $N_B$ fluctuates freely in each window. By simulating the $N$-th 
window, one immediately obtains the free energy difference
\beq
 \Delta F(N|\mu,\mu_B) \equiv W(N) - W(N-1) = \ln \left( C_+ / C_- \right),
\eeq
with $C_+$ ($C_-$) the number of times that the (unbiased) simulation was in a 
state with $N$ ($N-1$) particles of type $A$ (irrespective of $N_B$). Obviously, 
the free energy difference depends on $\mu$ and $\mu_B$. Once the free energy 
differences have been measured over a range of windows, $W(N)$ can be 
constructed via recursion
\beq
 W(0) \equiv 0, \hspace{5mm} W(N) = W(N-1) + \Delta F(N).
\eeq
Using that $P(N) \propto e^{W(N)}$, one trivially converts to the sought-for 
distribution $P(N|\mu,\mu_B)$.

An additional ingredient of this work is histogram reweighting, which we use to 
extrapolate simulation data obtained at $(\mu,\mu_B)$ to different chemical 
potentials $(\mu',\mu_B')$. In order to extrapolate in $\mu_B$, we also require 
the distributions $R(N_B|N,\mu_B)$, defined as the probability to observe a 
system containing $N_B$ particles of type $B$, when the number of $A$ particles 
equals $N$, at chemical potential $\mu_B$ (since $R(N_B|N,\mu_B)$ is obtained 
for fixed $N$, there is no dependence on $\mu$). For example, $R(N_B|0,\mu_B)$ 
is the distribution in $B$ particles when no $A$ particles are present (this 
corresponds to an ideal gas at chemical potential $\mu_B$, and hence a single 
Poissonian peak). The expression to extrapolate the free energy difference 
obtained at $(\mu,\mu_B)$ to $(\mu',\mu_B')$ then becomes
\begin{eqnarray}\label{eq:ext}
 \Delta F(N|\mu',\mu_B') = \hspace{4cm} \nonumber \\ 
 \Delta F(N|\mu,\mu_B) + \mu' - \mu + \ln \frac{ {\cal Z}(N) }{ {\cal Z}(N-1) },
\end{eqnarray}
with 
\beq\label{eq:zz}
 {\cal Z}(N) = \frac{ \sum_{N_B} R(N_B|N,\mu_B) e^{(\mu_B' - \mu_B) N_B} }{
 \sum_{N_B} R(N_B|N,\mu_B) }.
\eeq
Note that ${\cal Z}(N)$ is simply the relative change in the \ahum{volume} of 
$R(N_B|N,\mu_B)$ when extrapolating from $\mu_B \to \mu_B'$. By using 
\eq{eq:ext}, in combination with the recursion relation, it becomes possible to 
construct $P(N|\mu',\mu_B')$, without actually having to perform a simulation at 
$\mu'$ and $\mu_B'$.

The quality of \eq{eq:ext} deteriorates when the range in chemical potential 
over which one extrapolates becomes large. For each system size, we therefore 
perform a series of $i=1,\ldots,k$ SUS simulations, over a range of chemical 
potentials $\mu_i$ and $\mu_{B,i}$. Due to symmetry, we set $\mu_i = \mu_{B,i}$ 
for convenience, although this is not essential. Using \eq{eq:ext}, each one of 
these simulations yields an estimate of the free energy difference $\Delta 
F_i(N|\mu,\mu_B) \pm \sigma_i$, where $\sigma_i$ is an estimate of the 
statistical error. Next, we weight each estimate with its inverse square error 
to obtain the best estimate of the free energy difference
\beq
 \Delta F_{\rm best}(N|\mu,\mu_B) = \frac{ \sum_{i=1}^k 
 \Delta F_i (N|\mu,\mu_B) / \sigma_i^2 }{ \sum_{i=1}^k 1/\sigma_i^2 },
\eeq
which is then fed into the recursion relation to construct the best estimate of 
$P(N|\mu,\mu_B)$ for some $\mu,\mu_B$ of interest. To derive $\sigma_i$, we note 
that statistical errors occur in the counts $C_+$ and $C_-$, as well as in the 
histogram entries $R(N_B|N,\mu_B)$. In principle, these errors are Poissonian
\begin{eqnarray*}
 \sigma[C_+] = \sqrt{C_+}, \hspace{5mm} 
 \sigma[C_-] = \sqrt{C_-}, \\
 \sigma \left[ R(N_B|N,\mu_B) \right] = \sqrt{R(N_B|N,\mu_B)},
\end{eqnarray*}
but only if the data are normalized to the number of independent measurements. 
This requires knowledge of the correlation time $\tau$, which is computationally 
expensive to obtain, since the acceptance rate of the cluster move, and hence 
$\tau$, depend sensitively on density, composition, and system size. Instead, we 
follow a more pragmatic approach, whereby $C_+$ and $C_-$ are normalized to the 
number of accepted cluster moves in the window. Similarly, the histogram 
$R(N_B|N,\mu_B)$ is normalized to the number of cluster moves which resulted in 
a state with $N$ particles of type $A$ {\it and} which involved a change in the 
number of $B$ particles. With these choices, we assume that the statistical 
errors become Poissonian, i.e.~proportional to square-roots, with a common 
proportionality constant. Next, a propagation of errors calculation can be 
applied to \eq{eq:ext} and \eq{eq:zz} to derive $\sigma_i$. A final important 
optimization is to combine the set of histograms $R_i(N_B|N,\mu_{B,i})$ from the 
various SUS simulations into one best estimate using the multiple histogram 
method \cite{ferrenberg.swendsen:1989, newman.barkema:1999}, and to subsequently 
use this best estimate to calculate ${\cal Z}(N)$ of \eq{eq:zz}.


\section{Results}

\begin{figure}
\begin{center}
\includegraphics[width=\figwidth]{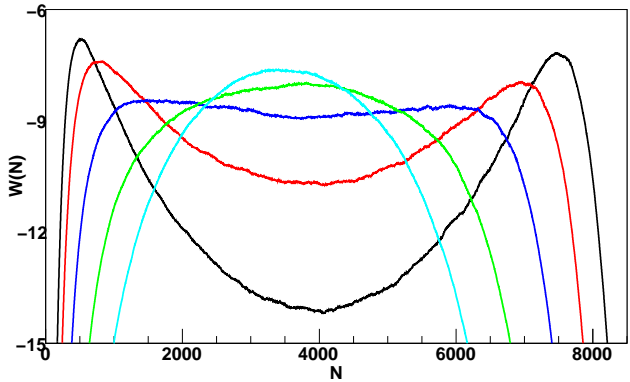} 

\caption{\label{wn} Distributions $W(N)$ obtained in a $40 \times 40$ system. 
The distributions were measured at $z_B=5.0, 5.1, 5.2, 5.3, 5.4$, and clearly 
illustrate the formation of the double-peaked structure with increasing $z_B$.}

\end{center}
\end{figure}

\begin{figure}
\begin{center}
\includegraphics[width=\figwidth]{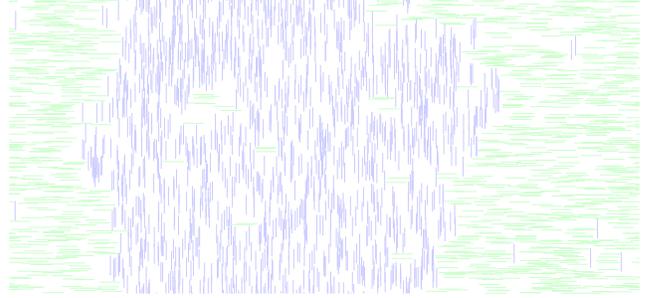} 
\caption{\label{snap} Snapshot of the 2D Zwanzig model at coexistence in a 
rectangular $L \times D$ simulation box, with $L=30$ and $D=15$, at fugacity 
$z_B=6.0$.}
\end{center}
\end{figure}

In \fig{wn}, we show distributions $W(N) = \ln P(N)$, for a number of fugacities 
$z_B$. For sufficiently large $z_B$, the distributions develop two pronounced 
peaks: one at low density $\rho=N/V$ of $A$ particles, and one at high density. 
Note that the bimodal structure only shows-up if the fugacity $z$ of the $A$ 
particles is chosen suitably. For the present model, of course, we may set 
$z=z_B$ due to symmetry, which was adopted throughout this work. For asymmetric 
fluids, choosing $z$ is less straightforward, and many criteria can, in fact, be 
defined \cite{orkoulas.fisher.ea:2001}.

\begin{figure}
\begin{center}
\includegraphics[width=\figwidth]{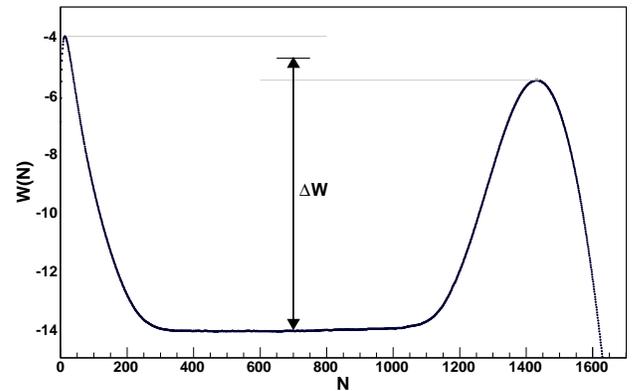} 
\caption{\label{wnflat} $W(N)$ as obtained in a rectangular $25 \times 10$ 
system at $z_B=6.0$. Note the flat region in between the peaks, and also the 
definition of the barrier $\Delta W$.}
\end{center}
\end{figure}

The connection to the liquid-gas transition becomes clear if one 
\ahum{identifies} the low-density peak with the gas phase, the high-density peak 
with the liquid, and $z_B$ with inverse temperature. The region between the 
peaks reflects phase coexistence, whereby both phases appear simultaneously. In 
simulations, the coexistence can be visualized directly (\fig{snap}). Note that 
a rectangular simulation box is used, and so the interfaces form parallel to the 
short edge, since this minimizes the total amount of interface (due to periodic 
boundary conditions, two interfaces are actually present). The corresponding 
distribution $W(N)$ for the rectangular system is shown in \fig{wnflat}. Note 
the flat region between the peaks, implying that interactions between the 
interfaces are absent. Following Binder \cite{binder:1982}, the height of the 
barrier $\Delta W$ in \fig{wnflat} yields the line tension $\sigma_l = \Delta W 
/ 2D$, with $D$ the short edge of the rectangle. For $z_B=6.0$ we obtain 
$\sigma_l \approx 0.47$ (in units of $k_B T$ per particle length). As expected, 
the line tension increases rapidly with increasing $z_B$, as manifested by the 
growing barriers of \fig{wn}.

\begin{figure}
\begin{center}
\includegraphics[width=\figwidth]{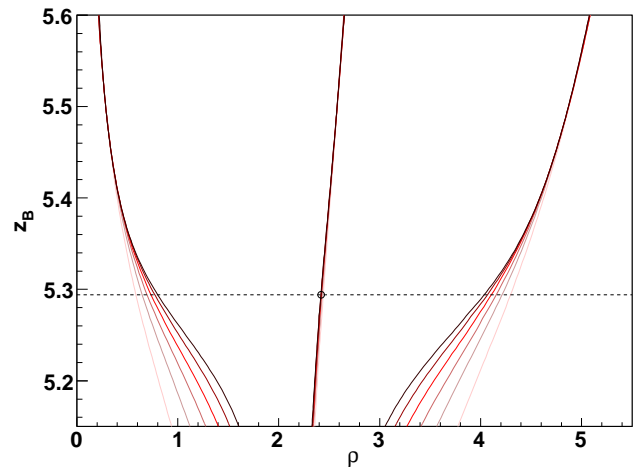}
\caption{\label{pd} Phase diagram (binodal) of the 2D Zwanzig model for system 
sizes $L=15,20,\dots,40$ (from outer to inner). The left (right) branches mark 
the positions of the gas (liquid) peak, the horizontal line is the critical 
fugacity $\zbcr \approx 5.294$ obtained from \fig{bc}(a). Also shown are the diameters 
(middle curves). The intersection of the diameter with the horizontal line 
yields the critical point (circle).}
\end{center}
\end{figure}

\begin{figure}
\begin{center}
\includegraphics[width=\figwidth]{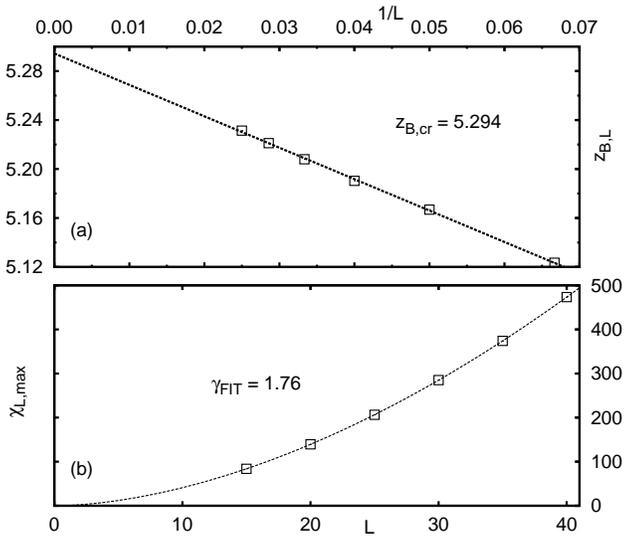}

\caption{\label{bc} Finite size scaling analysis of the susceptibility $\chi_L$. 
(a) Positions of the peak maxima $z_{B,L}$ versus $1/L$. The line is a linear fit 
from which $\zbcr$ follows. (b) The values $\chi_{L,\rm max}$ at the maxima versus 
$L$. The curve is a fit to a powerlaw, see details in the text, from which the 
critical exponent $\gamma$ follows.}

\end{center}
\end{figure}

In analogy to the liquid-gas transition, we can construct a binodal, by 
plotting the peak positions in $P(N)$ as a function of $z_B$ (\fig{pd}). 
In the thermodynamic limit $L \to \infty$, the gas and liquid branches of 
the binodal meet at the critical point, while in finite systems they 
\ahum{sway} around it (finite-size rounding \cite{binder.heermann:2002}). 
The horizontal line in \fig{pd} marks the thermodynamic limit estimate of 
the critical fugacity $\zbcr$, taken from \fig{bc}. Also shown in \fig{pd} 
are the \ahum{diameters}, defined as the average of the gas and liquid 
peak positions. In contrast to the binodal, finite size effects in the 
diameter are much weaker, the reason being that the singularity in the 
latter is only logarithmic in two dimensional Ising systems 
\cite{vink:010102}. From the intersection of the diameter with the line 
$\zbcr$ in \fig{pd}, we obtain $\rho_{\rm cr} \approx 2.42$ for the 
critical density of $A$ particles. Due to symmetry, the {\it overall} 
particle density is twice this value.

The critical fugacity $\zbcr$ was obtained from a scaling analysis of the 
finite-size susceptibility \cite{orkoulas.fisher.ea:2001}
\beq
 \chi_L = \frac{\avg{m^2} - \avg{|m|}^2}{L^2},
\eeq
with $m = N - \avg{N}$. Following standard arguments \cite{newman.barkema:1999}, 
$\chi_L$ versus $z_B$ exhibits a maximum, at position $z_{B,L}$ and with value 
$\chi_{L,\rm max}$. The peak positions scale with $L$ according to $z_{B,L} - 
\zbcr \propto L^{-1/\nu}$, with $\nu$ the critical exponent of the correlation 
length. Using the 2D Ising value $\nu=1$, we observe excellent scaling, see 
\fig{bc}(a), and by fitting we obtain $\zbcr \approx 5.294$. Of course, due to 
symmetry, the critical fugacity of the $A$ particles is also equal to this value. 
In addition, the peak maxima are expected to scale as $\chi_{L,\rm max} \propto 
L^{\gamma/\nu}$, with $\gamma$ the critical exponent of the susceptibility. 
Indeed, the maxima scale accordingly, see \fig{bc}(b), and by fitting we obtain 
$\gamma \approx 1.76$, in good agreement with the 2D Ising value $\gamma_{2D,I} = 
7/4$.

\section{Discussion}

We have shown that the 2D Zwanzig model undergoes a liquid-gas transition at 
critical fugacity $\zcr \approx 5.294$. As expected for systems with 
short-ranged interactions, the transition belongs to the universality class of 
the 2D Ising model. Note that the model studied here corresponds to that of hard 
rods on square lattices, in the limit where the rod length $k \to \infty$. As 
was shown in \cite{matoz-fernandez.linares.ea:2008}, the square lattice variant 
also exhibits a 2D Ising critical point, consistent with our findings. In these 
and other works \cite{ghosh.dhar:2007, matoz-fernandez.linares.ea:2008, 
linares.roma.ea:2008, matoz-fernandez:214902}, the resulting order above $\zcr$ 
is termed {\it nematic}, and the corresponding transition at $\zcr$ an 
isotropic-to-nematic transition. In contrast, our work indicates that the 
transition is just the liquid-gas transition. Therefore, the resulting order 
should be termed {\it magnetic}, since the liquid-gas transition is isomorphic 
to the formation of a spontaneous magnetization in the Ising model below its 
critical temperature.

Further evidence in favor of a liquid-gas transition, and against 
isotropic-to-nematic, is the coexistence between ordered states, see \fig{snap}. 
Note that phase coexistence in the 2D Zwanzig model is possible for 
all fugacities in the ordered region $z > \zcr$ of the phase diagram. The latter 
is analogous to the coexistence between domains of negative and positive 
magnetization in the Ising model below its critical temperature. In both cases, 
the line tension increases as one moves deeper into the ordered region. The 
corresponding interfaces are therefore {\it order-order} interfaces, which 
(likely) do not exist between nematic phases of liquid crystals, where the 
particle orientations are continuous \cite{citeulike:3558361, 
citeulike:3535124}. Of course, liquid crystals may exhibit isotropic-nematic 
coexistence, at a first-order isotropic-to-nematic transition. In that case one 
has {\it order-disorder} interfaces, which survive only {\it at} the transition 
point.

For the 2D Widom-Rowlinson (WR) model consisting of disks, the (single-species) 
critical density equals $\rho_{\rm cr} \approx 0.78$, while for the critical 
fugacity $\zcr \approx 1.73$ is obtained \cite{johnson.gould.ea:1997}. These 
values are significantly below the Zwanzig values reported here, implying a 
rather weak depletion effect in the latter. This can be made plausible by 
considering the excluded volume per particle. In a fluid of line segments, 
unlike segments exclude a volume $l^2$, with $l$ being the length. In a mixture 
of disks, one obtains $\pi l^2$, with $l$ being the disk diameter. Hence, the 
excluded volume per particle is $\pi$ times larger in fluids consisting of 
disks, and so we expect the transition at a density and fugacity reduced by 
roughly the same factor. Indeed, inspection of the reported numerical estimates 
follow this prediction reasonably well (to within 3\%).

Our main conclusion is therefore that phase transitions observed in the 2D 
Zwanzig model and its lattice variants should not be compared to liquid crystal 
transitions, but instead to transitions observed in simple fluids. This not only 
includes liquid-gas transitions, but also the possibility of crystallization at 
high density. Of course, the model considered in the present work cannot 
crystallize, since the particles are infinitely thin, but the lattice variants 
considered elsewhere may \cite{ghosh.dhar:2007, matoz-fernandez.linares.ea:2008, 
linares.roma.ea:2008, matoz-fernandez:214902}. Interestingly, these works indeed 
report evidence of a second transition occurring at high density; whether this 
transition can be interpreted in terms of (quasi) crystallization 
\cite{revmodphys.60.161} could be an interesting topic.

Finally, we discuss the expected trends when the 2D Zwanzig model is extended to 
a larger set of particle orientations. For hard rods on {\it triangular} 
lattices, the universality class is that of the 2D $3$-state Potts model 
\cite{matoz-fernandez.linares.ea:2008}. We expect the same universality class 
for an {\it off-lattice} fluid of hard line segments, with three allowed 
orientations $\theta \in \{ 0,\pi/3,2\pi/3 \}$ per segment (with $\theta$ the 
angle between, say, the segment and the $x$-axes). In the Potts model 
\cite{revmodphys.54.235}, the nearest-neighbor spin interaction assumes two 
values: a low (energetically favorable) value when two neighboring spins are in 
the same state, and a higher value when they are not. For line segments, the 
analogue is the excluded volume $a_d=|\sin \phi|$ between pairs of segments, 
with $\phi$ the angle between the segments. The reader can verify that for the 
above set of three orientations, one either has $a_d=0$ (when two segments are 
aligned) or $a_d=\sqrt{3}/2$ (when they are not). The excluded volume term thus 
exhibits the same symmetry as the Potts pair interaction, which naturally 
accounts for the observed critical behavior on triangular lattices 
\cite{matoz-fernandez.linares.ea:2008}. However, for larger sets of allowed 
orientations, we expect the analogy to the Potts model to break down. For 
instance, allowing four orientations $\theta \in \{ 0, \pi/4, \pi/2, 3\pi/4 \}$, 
the excluded volume already assumes three values, in disagreement with Potts 
symmetry. Note that the analogy to the Potts model may also be useful to 
interpret results obtained in three dimensions. In this case, the Zwanzig model 
consists of hard rods of finite width, allowed to point in three mutually 
perpendicular directions. In terms of symmetry, this corresponds to a 
three-dimensional $3$-state Potts model, which has a first-order phase 
transition \cite{citeulike:3574310, citeulike:3683252}. The latter is indeed 
consistent with Zwanzig's result, where the existence of a first-order 
transition was also demonstrated \cite{zwanzig:1714}. In three dimensions, it 
should also be possible to realize $4$-state Potts symmetry, by choosing the rod 
orientations perpendicular to the faces of a regular tetrahedron. For larger 
sets of orientations, the analogy to the Potts model is again expected to break 
down. We refer the interested reader to \cite{citeulike:3574333, 
shundyak.roij:2004} where Zwanzig models with large sets of allowed particle 
orientations are discussed.

In summary, we have presented simulation results of the 2D Zwanzig model. In 
agreement with lattice variants of this model, we find a phase transition with 
critical point belonging to the universality class of the 2D Ising model. The 
novelty of the present work has been the interpretation of this transition in 
terms of the liquid-gas transition, as opposed to isotropic-to-nematic. This 
interpretation accounts naturally for the observed critical behavior, as well as 
for the observed phase coexistence in the ordered region of the phase diagram. 
In addition, the excluded volume interaction in fluids of hard line segments 
with restricted orientations, was shown to resemble in some cases the symmetry 
of the $q$-state Potts model with $q>2$, thereby elucidating a previous 
simulation result \cite{matoz-fernandez.linares.ea:2008}.

\acknowledgments This work was supported by the {\it Deutsche 
Forschungsgemeinschaft} under the Emmy Noether program (VI~483/1-1).

\bibliography{mc1975}

\end{document}